\documentclass[pdflatex,sn-chicago]{sn-jnl}

\usepackage{graphicx}%
\usepackage{multirow}%
\usepackage{amsmath,amssymb,amsfonts}%
\usepackage{amsthm}%
\usepackage{mathrsfs}%
\usepackage[title]{appendix}%
\usepackage{xcolor}%
\usepackage{textcomp}%
\usepackage{manyfoot}%
\usepackage{booktabs}%
\usepackage{algorithm}%
\usepackage{algorithmicx}%
\usepackage{algpseudocode}%
\usepackage{listings}%

\usepackage{rotfloat}

\def\BibTeX{{\rm B\kern-.05em{\sc i\kern-.025em b}\kern-.08em
    T\kern-.1667em\lower.7ex\hbox{E}\kern-.125emX}}

\usepackage[nopostdot,style=super,nonumberlist,toc]{glossaries}
\glsdisablehyper

\makeglossaries
\newacronym{VM}{VM}{Virtual Machine}
\newacronym{HPC}{HPC}{High Performance Computing}
\newacronym{SLA}{SLA}{Service-Level Agreement}
\newacronym{DTMC}{DTMC}{Discrete-Time Markov Chain}
\newacronym{MPI}{MPI}{Message Passing Interface}
\newacronym{BLCR}{BLCR}{Berkeley Lab Checkpoint/Restart}
\newacronym{NVCR}{NVCR}{NVIDIA Checkpoint/Restart}
\newacronym{DMTCP}{DMTCP}{Distributed MultiThreaded CheckPointing}
\newacronym{CRIU}{CRIU}{Checkpoint/Restore In Userspace}
\newacronym{P.Haul}{P.Haul}{Process Hauler}
\newacronym{FTI}{FTI}{Fault Tolerance Interface}
\newacronym{C/R}{C/R}{Checkpoint/Restore}
\newacronym{RPC}{RPC}{Remote Procedure Call}
\newacronym{PID}{PID}{Process Identifier}
\newacronym{RDMA}{RDMA}{Remote Direct Memory Access}
\newacronym{MTC}{MTC}{Migration Toolkit for Containers}
\newacronym{JNLP}{JNLP}{Joint Network and Local Prediction}
\newacronym{DCL}{DCL}{Dynamic Context Loading}
\newacronym{LCR}{LCR}{Lightweight Container Registry}
\newacronym{KVM}{KVM}{Kernel-based Virtual Machine}
\newacronym{RWX}{RWX}{ReadWriteMany}
\newacronym{MTV}{MTV}{Migration Toolkit for Virtualization}
\newacronym{RHV}{RHV}{RedHat Virtualization}
\newacronym{HCI}{HCI}{Hyper-Converged Infrastructure}
\newacronym{QEMU}{QEMU}{Quick Emulator}
\newacronym{IoT}{IoT}{Internet of Things}
\newacronym{LXD}{LXD}{Linux Container Daemon}
\newacronym{LXC}{LXC}{Linux Container}
\newacronym{HETCOM}{HETCOM}{Heterogeneous Container Migration}
\newacronym{ISA}{ISA}{Instruction Set Architectures}
\newacronym{IR}{IR}{Intermediate Representation}
\newacronym{TEE}{TEE}{Trusted Execution Environment}
\newacronym{TPM}{TPM}{Trusted Platform Module}
\newacronym{MANA}{MANA}{MPI-Agnostic Network-Agnostic Transparent Checkpointing}
\newacronym{K8s}{K8s}{Kubernetes}
\newacronym{HA}{HA}{High Availability}
\newacronym{VLAN}{VLAN}{Virtual Local Area Network}
\newacronym{PV}{PV}{paravirtualization}
\newacronym{HVM}{HVM}{Hardware Virtualized Machines}
\newacronym{VE}{VE}{Virtual Env.}
\newacronym{LAN}{LAN}{Local Area Network}
\newacronym{WAN}{WAN}{Wide Area Network}
\newacronym{GPU-P}{GPU-P}{GPU partitioning}
\newacronym{CNF}{CNF}{Cloud-Native Network Functions}
\newacronym{LIMOCE}{LIMOCE}{Live Migration of Containers in the Edge}
\newacronym{rootfs}{rootfs}{root filesystem}
\newacronym{CGroups}{CGroups}{Control Groups}
\newacronym{IaaS}{IaaS}{Infrastructure-as-a-Service}
\newacronym{POS}{POS}{PARALLELGPUOS}
\newacronym{NPT}{NPT}{Nested Page Table}
\newacronym{multifd}{multifd}{multiple file descriptors} 

\begin{document}

\title[Seamless Transitions: A Comprehensive Review of Live Migration Technologies]{Seamless Transitions: A Comprehensive Review of Live Migration Technologies}

\author*[1,2]{\fnm{Sima} \sur{Attar-Khorasani}}\email{sima.attar\_khorasani@tu-dresden.de}

\author[1,2]{\fnm{Lincoln} \sur{Sherpa}}\email{lincoln.sherpa@tu-dresden.de}

\author[1,2]{\fnm{Matthias} \sur{Lieber}}\email{matthias.lieber@tu-dresden.de}

\author[2,3]{\fnm{Siavash} \sur{Ghiasvand}}\email{siavash.ghiasvand@tu-dresden.de}

\affil[1]{\orgdiv{Information Services and High Performance Computing} \orgname{(ZIH)}, \orgname{TUD Dresden University of Technology}, \orgaddress{\city{Dresden}, \country{Germany}}}

\affil[2]{\orgdiv{Center for Interdisciplinary Digital Sciences (CIDS)}, \orgname{TUD Dresden University of Technology}, \orgaddress{\city{Dresden}, \country{Germany}}}

\affil[3]{\orgdiv{Center for Scalable Data Analytics and Artificial Intelligence}, \orgname{(ScaDS.AI)}, \orgaddress{\city{Dresden/Leipzig}, \country{Germany}}}

\abstract{Live migration, a technology enabling seamless transition of operational computational entities between various hosts while preserving continuous functionality and client connectivity, has been the subject of extensive research.
However, existing reviews often overlook critical technical aspects and practical challenges integral to the usage of live migration techniques in real-world scenarios.
This work bridges this gap by integrating the aspects explored in existing reviews together with a comprehensive analysis of live migration technologies across multiple dimensions, with focus on migration techniques, migration units, and infrastructure characteristics.
Despite efforts to make live migration widely accessible, its reliance on multiple system factors can create challenges.
In certain cases, the complexities and resource demands outweigh the benefits, making its implementation hard to justify.
The focus of this work is mainly on container-based and virtual machine-based migration technologies, examining the current state of the art and the disparity in adoption between these two approaches.
Furthermore, this work explores the impact of migration objectives and operational constraints on the usability and efficacy of existing technologies.
By outlining current technical challenges and providing guidelines for future research and development directions, this work serves a dual purpose: first, to equip enthusiasts with a valuable resource on live migration, and second, to contribute to the advancement of live migration technologies and their practical implementation across diverse computing environments.}

\keywords{Live Migration, Container, Virtual Machine, Live Migration Survey, Checkpoint/Restore Tools}

\maketitle



\section{Introduction} 
Live migration denotes seamless transfer of a running computational entity, such as a \gls{VM}, between various hosts while the continuous operation and client connectivity is maintained.
First demonstrated in 2005~\cite{Clark2005vm}, live migration has evolved to encompass applications in Edge, cloud, and \gls{HPC}, offering various benefits in flexibility, scalability, fault tolerance, and resource efficiency.

Various studies reviewed live migration.
However, most reviews focus primarily on categorizing existing techniques and comparing their performance metrics, without delving into the fundamental technical aspects of live migration.
Others either overlook the practical challenges in performing live migration, or they focus heavily on particular applications or environments.
Therefore, a significant gap exists in studying live migration technologies and real-world operational scenarios constrained by practical considerations.
This work addresses this gap by analyzing existing technologies, exploring unique challenges, and examining migration specifications tailored to diverse practical perspectives, focusing on four key dimensions namely migration technique, migration unit, infrastructure characteristics, and migration objective.
The various perspectives and decision criteria are further summarized in a table structure to make the complexities and relations of live migration technologies better understandable, providing insights into how live migration solutions can be adapted to meet the distinct needs in various scenarios.

Among the four key dimensions in this work, the \textit{migration technique} encompasses various methods, such as pre-copy, post-copy, and hybrid.
The \textit{migration unit} investigates the differences between containers and \gls{VM}s, each with distinct requirements and technologies in the migration context.
The \textit{infrastructure characteristics} explores how factors such as network conditions and environment influence migration strategies.
The \textit{migration objective} examines the trade-offs involved, such as minimizing downtime, optimizing resource usage, and addressing complexities like debugging or system intricacies.
It is also crucial to note that underlying technologies, such as checkpointing, play a vital role in the live migration process.
Such variations significantly influence the efficiency and effectiveness of live migration across different contexts.

This work provides researchers with a clear understanding of the current state-of-the-art in live migration, highlighting areas where current technologies excel and identifying gaps requiring further innovation.
This work is organized as follows:
Section~\ref{sec:overview} presents a comprehensive overview of relevant studies and defines the scope of this research.
Sections~\ref{sec:techniques},~\ref{sec:units}, and~\ref{sec:characteristics} investigate the effects of migration techniques, units, and characteristics on live migration, respectively.
To conclude, Section~\ref{sec:discussion} examines the significance of migration objectives and provides practical guidelines for live migration implementation.
\section{Scope of the work and relevant studies}
\label{sec:overview}
The ISO 13008 standard defines the term migration as the transfer of records from one hardware or software configuration to another, ensuring the preservation of their original format through a well-defined and structured set of principles~\cite{iso13008}.
After migration, the  service or software component is required to comply with precise specifications and rigorous quality evaluation criteria to ensure optimal performance and reliability\footnote{For example, the ISO 25010 and ISO 25011 standards~\cite{iso25010,iso25011}.}.
These criteria play a vital role in guaranteeing that the migrated entity maintains its functionality, performance, and reliability within the new environment, ensuring the integrity and usability of the transferred records or applications remain intact.
In the context of process migration, three primary categories are typically identified: no migration, cold migration, and live migration.
In the first category, the process remains on its original host, while in the subsequent two categories the process will be moved to another host.
Live migration expands the concept of traditional migration by enabling the seamless transition of a running computational entity, such as a \gls{VM} or container, from one host to another with little to no service disruption.
This stands in contrast to cold migration, which allows service disruption during the migration from the source to the destination.

Live migration performance is measured by (1) migration time -- the total duration from the start to the end of the migration process -- and (2) migration downtime -- the period of service unavailability~\cite{performance_driven_VM}.
The length of downtime determines whether migration is classified as live or cold, depending on its acceptability for a given use case.
Classifying service interruption is also highly context-dependent as it varies according to individual use cases, technical intricacies, and performance metrics unique to each scenario.
As a result, creating a universal definition of live migration with fixed, standardized metrics that applies to all situations is impractical
~\cite{live_migration_def, intelligent_docker_container}.

\begin{sidewaystable}[htbp]
\caption{An overview of recent studies on live migration. $\bullet$ denotes broad discussion. $\circ$ denotes partial discussion.}\label{tab:recent_surveys}
\setlength{\tabcolsep}{4.25pt}
\begin{tabular*}{\textwidth}{@{}lcccccccccccccccccccccccccccc}

\toprule%
~&
\multicolumn{3}{@{}c@{}}{\centering\parbox{1.6cm}{\centering Migration\\Issues}}~&
\multicolumn{4}{@{}c@{}}{\centering\parbox{1.6cm}{\centering Migration\\Units}}  &
\multicolumn{2}{@{}c@{}}{\centering\parbox{1.6cm}{\centering Env.}} &
\multicolumn{3}{@{}c@{}}{\centering\parbox{1.6cm}{\centering Migration\\Technology}}  &
\multicolumn{3}{@{}c@{}}{\centering\parbox{1.6cm}{\centering C/R\\Tools}}  &
\multicolumn{5}{@{}c@{}}{\centering\parbox{1.6cm}{\centering Container\\Engines}}  &
\multicolumn{7}{@{}c@{}}{\centering\parbox{1.6cm}{\centering Virtualization\\Tools}}  \\
\cmidrule{2-4}
\cmidrule(lr{0.2em}){5-8}
\cmidrule(lr{0.2em}){9-10}
\cmidrule(lr{0.2em}){11-13}
\cmidrule(lr{0.2em}){14-16}
\cmidrule(lr{0.2em}){17-21}
\cmidrule{22-28}

        ~ &
        \rotatebox[origin=r]{270}{Mechanism}     & \rotatebox[origin=r]{270}{Application}          & \rotatebox[origin=r]{270}{Management} &
        \rotatebox[origin=r]{270}{VM}       & \rotatebox[origin=r]{270}{Container}            & \rotatebox[origin=r]{270}{Network} &
        \rotatebox[origin=r]{270}{Storage}        & \rotatebox[origin=r]{270}{Cloud/Cluster}        &  \rotatebox[origin=r]{270}{Edge/Fog} & 
        \rotatebox[origin=r]{270}{Pre-Copy}        & \rotatebox[origin=r]{270}{Post-Copy}             &  \rotatebox[origin=r]{270}{Hybrid Copy} &
        \rotatebox[origin=r]{270}{CRIU}     &  \rotatebox[origin=r]{270}{DMTCP} &
        \rotatebox[origin=r]{270}{Others\footnotemark[1]}& \rotatebox[origin=r]{270}{Docker}              & \rotatebox[origin=r]{270}{LXC} &
        \rotatebox[origin=r]{270}{OpenVZ}         & \rotatebox[origin=r]{270}{Apptainer} & \rotatebox[origin=r]{270}{Podman}    & \rotatebox[origin=r]{270}{KVM/QEMU } &
        \rotatebox[origin=r]{270}{VMware}         & \rotatebox[origin=r]{270}{Hyper-V}             & \rotatebox[origin=r]{270}{Xen} &
        \rotatebox[origin=r]{270}{VirtualBox}     & \rotatebox[origin=r]{270}{Virtuozzo}           & \rotatebox[origin=r]{270}{Others\footnotemark[2]}   \\ 
\midrule

\cite{Strunk2012costs}~&~$\bullet$~&~&~&~$\bullet$~&~&~&~&~$\bullet$~&~&~$\bullet$~&~&~&~&~&~&~&~&~&~&~&~&~&~&~&~&~&~\\ 
\cite{Shetty2012Secure}~&~$\bullet$~&~&~&~$\bullet$~&~&~$\bullet$~&~&~&~&~&~&~&~&~&~&~&~&~&~&~&~&~&~&~&~&~&~\\ 
\cite{Xu2014Performance}~&~$\bullet$~&~&~&~$\bullet$~&~&~&~&~$\bullet$~&~&~&~&~&~&~&~&~&~&~&~&~&~&~&~&~&~&~&~\\ 
\cite{Medina2014survey}~&~$\bullet$~&~$\circ$~&~&~$\bullet$~&~$\circ$~&~$\circ$~&~$\circ$~&~$\circ$~&~&~$\bullet$~&~$\bullet$~&~&~&~&~&~&~&~$\bullet$~&~&~&~$\bullet$~&~$\bullet$~&~&~$\bullet$~&~&~&~\\ 
\cite{Li2015comparing}~&~$\bullet$~&~&~&~$\bullet$~&~$\bullet$~&~&~&~$\bullet$~&~&~&~&~&~$\bullet$~&~&~&~$\bullet$~&~$\circ$~&~$\bullet$~&~&~&~&~$\bullet$~&~&~$\bullet$~&~&~&~\\ 
\cite{Yamada2016Survey}~&~$\bullet$~&~&~&~$\bullet$~&~&~$\circ$~&~$\circ$~&~&~&~$\bullet$~&~$\bullet$~&~$\bullet$~&~&~&~&~&~&~&~&~&~$\circ$~&~&~&~$\circ$~&~&~&~\\
\cite{Zhang2018network}~&~$\bullet$~&~$\bullet$~&~&~$\bullet$~&~$\circ$~&~$\bullet$~&~$\bullet$~&~$\bullet$~&~$\bullet$~&~$\bullet$~&~$\bullet$~&~$\bullet$~&~&~&~&~&~&~&~&~&~&~&~&~&~&~&~\\
\cite{Wang2018Survey}~~&~$\circ$~&~$\bullet$~&~$\circ$~&~$\bullet$~&~$\bullet$~&~$\circ$~&~&~&~$\bullet$~&~$\circ$~&~$\circ$~&~&~&~&~&~&~&~&~&~&~&~&~&~&~&~&~\\
\cite{NOSHY2018optimization}~&~$\bullet$~&~&~&~$\bullet$~&~&~&~&~$\bullet$~&~&~$\bullet$~&~$\bullet$~&~$\bullet$~&~&~&~&~&~&~&~&~&~$\bullet$~&~&~&~$\bullet$~&~&~&~\\
\cite{Rejiba2019Survey}~&~&~$\bullet$~&~&~$\circ$~&~$\bullet$~&~$\bullet$~&~&~&~$\bullet$~&~$\circ$~&~$\circ$~&~$\circ$~&~&~&~&~&~&~&~&~&~&~&~&~&~&~&~\\
\cite{LE2020survey}~~~&~$\bullet$~&~&~&~$\bullet$~&~&~&~&~$\bullet$~&~&~$\bullet$~&~$\bullet$~&~$\bullet$~&~&~&~&~&~&~&~&~&~$\bullet$~&~$\bullet$~&~$\bullet$~&~$\bullet$~&~$\bullet$~&~$\bullet$~&~\\
\cite{masdari2020efficient}~&~$\bullet$~&~&~&~$\bullet$~&~&~~&~&~$\bullet$~&~&~$\bullet$~&~$\bullet$~&~$\bullet$~&~&~&~&~&~&~&~&~&~$\bullet$~&~&~&~$\bullet$~&~&~&~\\
\cite{Ramanathan2021vm}~&~$\bullet$~&~&~&~$\bullet$~&~$\bullet$~&~$\bullet$~&~&~$\bullet$~&~&~$\bullet$~&~&~&~$\bullet$~&~&~&~$\bullet$~&~&~&~&~&~$\bullet$~&~&~&~&~&~&~\\
\cite{imran2022live}~~&~&~$\circ$~&~$\bullet$~&~$\bullet$~&~$\circ$~&~$\bullet$~&~$\circ$~&~$\bullet$~&~&~$\bullet$~&~$\bullet$~&~$\bullet$~&~&~&~&~&~&~&~&~&~&~&~&~&~&~&~\\
\cite{Kaur2022cloud}~~&~$\bullet$~&~&~&~$\circ$~&~$\bullet$~&~$\bullet$~&~$\bullet$~&~$\bullet$~&~$\bullet$~&~$\bullet$~&~$\bullet$~&~$\bullet$~&~&~&~&~&~&~&~&~&~&~&~&~&~&~&~\\
\cite{elsaid2022virtual}~&~&~&~$\bullet$~&~$\bullet$~&~&~$\bullet$~&~&~$\bullet$~&~&~$\bullet$~&~&~&~&~&~&~&~&~&~&~&~$\bullet$~&~$\bullet$~&~$\bullet$~&~$\bullet$~&~&~&~\\
\cite{soma2023vm}~~~~&~&~$\bullet$~&~&~$\bullet$~&~$\bullet$~&~~&~&~$\bullet$~&~&~$\circ$~&~&~&~&~&~&~&~&~$\bullet$~&~&~&~&~&~&~&~&~&~\\
\cite{live_migration_data_management}~&~$\circ$~&~$\bullet$~&~$\bullet$~&~$\bullet$~&~$\bullet$~&~$\bullet$~&~$\circ$~&~$\bullet$~&~$\bullet$~&~$\bullet$~&~$\bullet$~&~$\bullet$~&~&~&~&~&~&~&~&~&~&~&~&~&~&~&~\\
\cite{survey_live_migration_container}~&~$\bullet$~&~$\bullet$~&~&~$\bullet$~&~$\bullet$~&~$\bullet$~&~&~$\bullet$~&~$\bullet$~&~$\bullet$~&~$\bullet$~&~$\bullet$~&~$\bullet$~&~&~&~&~&~&~&~&~&~&~&~&~&~&~\\\hline
\textbf{Our~work}~&~$\bullet$~&~$\bullet$~&~$\bullet$~&~$\bullet$~&~$\bullet$~&~$\bullet$~&~$\bullet$~&~$\bullet$~&~$\bullet$~&~$\bullet$~&~$\bullet$~&~$\bullet$~&~$\bullet$~&~$\bullet$~&~$\bullet$~&~$\bullet$~&~$\bullet$~&~$\bullet$~&~$\bullet$~&~$\bullet$~&~$\bullet$~&~$\bullet$~&~$\bullet$~&~$\bullet$~&~$\bullet$~&~$\bullet$~&~$\bullet$~ \\

\botrule
\end{tabular*}
\footnotetext[1]{Hyper-V \gls{C/R}, Wharf, RedHat OpenShift, Fedora \gls{C/R}, H-Container, \gls{HETCOM}, \gls{FTI}, \gls{P.Haul}, \gls{BLCR}}
\footnotetext[2]{OpenStack, RedHat OpenShift, Harvester}
\end{sidewaystable}


An overview of the existing literature on live migration is presented in the rest of this section and summarized in Table~\ref{tab:recent_surveys}.
Emphasis is placed on recent studies that examine live migration from technical standpoints.
\cite{Kaur2022cloud} proposes a taxonomy of techniques for live container migration, focusing on various methods such as pre-copy and post-copy migration, prediction models, and bandwidth optimization.
It explores container migration schemes across three computing layers: cloud, Fog, and Edge computing.
A similar study~\cite{live_migration_data_management} offers a comprehensive overview with multiple management-related categories and subcategories concerning live migration in Edge and cloud computing environments. 

From an energy optimization perspective, particularly for data centers in cloud environments,~\cite{soma2023vm} provides an overview of research conducted on both containers and virtual machines.
The authors summarize key studies on live migration in these environments, outlining individual methods and results.

On the other hand,~\cite{imran2022live} classifies live virtual machine migration techniques from multiple perspectives: load balancing, energy efficiency, \gls{SLA} awareness, and network-awareness.
They provide a thorough discussion of various approaches within these categories, including push-pull strategies, hybrid bio-inspired algorithms, AI-based methods like Backpropagation Artificial Neural Network for load balancing, energy-aware schemes such as Efficient Adaptive Migration Algorithm and \gls{DTMC}, and \gls{SLA}-aware techniques that utilize reinforcement learning and neural networks.
The study effectively surveys various live \gls{VM} migration methods and compares their performance across multiple metrics. 



A growing number of review papers, such as~\cite{Ramanathan2021vm} conduct comparative analysis of different migration methods in Fog computing, focusing on key metrics such as total migration time, downtime, and the overall volume of transferred data.




In summary, while the existing literature on live migration covers generic aspects mainly within the scope of cloud computing environments and management-centric perspectives, this work extends the scope beyond these and provides a comprehensive technical evaluation of the comparative advantages, disadvantages, and trade-offs among different technologies, offering detailed comparisons and in-depth analyses of live migration solutions.
Furthermore, given the rapid evolution of the live migration field, regular updates are necessary to incorporate the latest advancements in live migration tools and technologies.



\section{Migration Techniques}
\label{sec:techniques}


Live migration primarily relies on shared storage due to its efficiency in transferring only process memory and state between hosts, reducing migration time and network load, while enabling simplified management.
However, alternative methods like “shared nothing” live migration are also available to increase flexibility~\cite{survey_live_migration_container}.
In shared storage live migration, only \gls{VM} memory, CPU, device, and network states are transferred, while shared nothing migration requires transferring the entire \gls{VM} storage in addition to the memory and states.
Data migration across hosts can be achieved via two primary approaches, distinguished by the order in which key components are transferred.
The first approach prioritizes the transfer of the process's memory content before moving other elements, while the second method initially transfers the process state, followed by other components.
Each of these approaches offers distinct advantages and trade-offs depending on the specific use case and system requirements.



\subsection{Pre-copy}\label{subsec2}
The pre-copy policy begins by transferring all memory pages and required files to the destination host while the process remains operational.
Subsequently, it iteratively transfers the modified data (dirty pages) during each iteration until the amount of modified data becomes minimal or satisfies predefined termination conditions.
At this point, the process is paused, the remaining dirty pages and state information are migrated, and finally, the process is restored to continue operation on the destination host \cite{intelligent_docker_container, live_migration_data_management}.
Live migration via pre-copy introduces challenges, such as limited parallelism during data dumping, non-convergence of the pre-copy strategy, and insufficient parallelism in the recovery process.
Recent studies aim to address these challenges by employing enhanced page servers, memory activity detection, Auto-converge, and machine learning-based behavior predictions~\cite{precopy_def, haris2024optimizing}.

\subsection{Post-copy}
This approach transfers the process's state to the target host and resumes it there before transferring the memory.
Memory pages are then actively pushed from the source to the target, and any missing pages are fetched on demand.
Strategies such as ensuring that each page is transferred only once, significantly reduce the copy redundancy compared to pre-copy method~\cite{live_migration_data_management, post_copy_migration}.
The post-copy method however, imposes a high risk of data loss if a failure occurs before all memory pages are transferred.
Therefore, although post-copy introduces less overhead comparing to pre-copy, and different variations of post-copy such as demand paging (lazy copy), active pushing, pre-paging, and dynamic self-ballooning~\cite{post_copy_migration} have been proposed, almost no real-world computing system adopts post-copy migration \cite{Puliafito2018} due to its lower reliability.

Considering the advantages and disadvantages of both pre-copy and post-copy methods, various hybrid methods were proposed to combine the strength of both methods.
Hybrid methods aim to transfer memory via pre-copy, and process's state via post-copy, thus enhancing both reliability and efficiency~\cite{live_migration_data_management,survey_live_migration_container,hybrid_approach}.
When the characteristics of a workload are predictable, more advanced variations such as speculative method~\cite{lu2013hsg} can be employed to further improve the overall performance.

It is worth emphasizing that the boundaries of live migration can only be defined in combination with a concrete use case.
This context-dependent nature allows even simple methods like stop-and-copy to be considered viable for live migration in specific scenarios; for instance, in interactive applications where migration downtime falls below the required interaction timeout.\newline
\indent Migration can be performed at various levels of granularity, ranging from page-level (smaller granularity) to the migration of an entire node (larger granularity).
The selected level of granularity plays a critical role in determining the overall performance and efficiency of the migration process.
Larger granularity can distribute migration overheads more effectively, but it risks wasting fast memory capacity if the application does not fully utilize the migrated data.
On the other hand, smaller granularity provides finer control over the process, though it may lead to increased overhead due to the need for more frequent transfer.
Existing studies and empirical observations suggest that while finer granularity may provide more precise control, the current state of technology, tooling, and industry practices favors \gls{VM} and container-level migration as the most practical and beneficial approach.
This is due to several key factors, including portability and flexibility, resource efficiency, scalability and management, development and deployment speed, application isolation, ecosystem support, and modernization pathways.
Therefore, this work focuses on live migration strategies at the \gls{VM} and container level.

\subsection*{Checkpoint/Restore Tools}
To implement these migration techniques, various \gls{C/R} tools have been developed to capture the state of a running process (checkpointing) and later recreate it as a new running process (restoring), effectively duplicating the original process context.
\gls{C/R} tools can operate at different levels, namely: system-level (e.g., \gls{BLCR}) executed in kernel space, user-level (e.g., \gls{DMTCP}, \gls{CRIU}) executed in user space, and application-level (e.g., \gls{FTI}) requiring modifications to the application code \cite{Jia2024DCU-CHK, Benjaponpitak2020blcr}.
Each \gls{C/R} tool pursues distinct goals and techniques,
~and the execution environment imposes different constraints.
~The rest of this section explores the technical characteristics of some the most prominent \gls{C/R} tools, listed in Table~\ref{tab:C/R-tools}.

\begin{table}[h]
\caption{Overview of Checkpoint/Restore tools}\label{tab:C/R-tools}
\begin{tabular*}{\textwidth}{@{\extracolsep\fill}lllll}
\toprule
\gls{C/R} Tools  &  Migration Tech.   &  Execution Env.  &  Limitations   &  Use cases \\
\midrule
\rule{0pt}{2.5ex}\gls{BLCR}             & Pre-copy & Kernel space            & Linux specific        & \gls{HPC}                \\
\gls{DMTCP}            & Pre-copy & User space              & Partial state         & \gls{HPC}                \\
\gls{CRIU}             & Pre-copy/Post. & User space            & Linux specific        & Container          \\
\gls{P.Haul}           & Pre-copy/Post. & User space            & Overhead              & Container          \\
\gls{FTI}              & Pre-copy & App. code               & Code modification     & \gls{HPC}                \\
OpenShift        & Pre-copy/Post. & OpenShift env.        & High Granularity      & Container          \\
Fedora \gls{C/R}       & Pre-copy & User space              & Platform specific     & Process            \\
Hyper-V \gls{C/R}      & Pre-copy & Hypervisor              & Windows specific      & \gls{VM}              \\
CAST AI          & Pre-copy/Post. & Clouds                & Platform specific     & Cloud              \\
H-Container      & Pre-copy/Post. & User space            & Linux specific        & Container          \\
\gls{HETCOM}           & Pre-copy & User space              & Linux specific        & Security            \\
Wharf            & Pre-copy & User space              & Adoption              & Application         \\
\botrule
\end{tabular*}
\end{table}

\subsubsection{Berkeley Lab Checkpoint/Restart}
\gls{BLCR} was a system-level checkpoint/restart implementation for Linux clusters that is no longer in active maintenance.
It had a hybrid kernel/user implementation that provides checkpoint/restore functionality without requiring changes to application code.
Based on \gls{BLCR} and its extensions, various supporting tools such as a middle-ware for load balancing between the nodes had been proposed~\cite{Gerofi2010blcr}.
\gls{BLCR} targeted checkpointing of \gls{HPC} applications by integrating with various \gls{MPI} implementations like MVAPICH and Open\,MPI. However, due to \gls{BLCR}'s lack of support for SysV shared memory, these implementations were unable to continue supporting \gls{BLCR} \cite{Xu2023Mana}.


\subsubsection{Distributed MultiThreaded CheckPointing}

\gls{DMTCP} is a user-space tool for checkpoint/restart of multi-threaded, distributed applications without kernel modifications that also integrates with \gls{HPC} environments and handles a wide range of system artifacts~\cite{dmptc_checkpointing_article, dmptc_article,dmptc_hpc}.
To use \gls{DMTCP}, a process or container must be initiated with a predefined library linked from the start, which could be regarded as a limitation~\cite{Poggiani2024lm}.
Additionally, it intercepts the process's glibc/kernel calls to create a checkpoint, however it is sometimes unable to capture certain calls, such as inotify() which might lead to inconsistencies or data loss~\cite{Poggiani2024lm}.
Comparing the performance of \gls{BLCR} and \gls{DMTCP} ~\cite{azeem2023performance} concludes that \gls{BLCR} has a weaker performance in terms of time needed for checkpointing and restarting.
\gls{DMTCP} requires a separate plugin for each network interconnect, making it costly to maintain and not fully network-agnostic.
To address this, the open-source tool \gls{MANA} was proposed, using a split-process mechanism that
allows to checkpoint/restart only the part of the process not belonging to the MPI library while tearing down and reinitializing the MPI part, relieving this limitation~\cite{Xu2023Mana}.

\subsubsection{Checkpoint/Restore In Userspace}
Originally released in 2012, \gls{CRIU} relies on Linux kernel features available in kernels since version $3.11$.
It is mainly known for its effective container migration capabilities~\cite{live_migration_data_management}.
It has been integrated into different container engines such as OpenVZ, Podman, Borg (Google Container Engine), \gls{LXC}/\gls{LXD}, Docker and others.
\gls{CRIU} pauses the migrating process using ptrace(), which freezes it in place.
Then it gathers all necessary process data from existing kernel interfaces in userspace and writes them as a condensed binary image to disk.
Next, parasite code is injected via an \gls{RPC}, allowing \gls{CRIU} to swiftly write all process memory pages to disk.
The parasite code is then removed, leaving the process unaware it was ever injected.
The Page-Server mechanism in \gls{CRIU} reduces filesystem overhead during migration by directly transferring a process's memory content from the source node to the destination node.
Instead of storing dump files locally and then transferring them, the Page-Server, running on the destination node, can receive the memory data directly, minimizing storage needs on the source node \cite{criu_hpc_application}. 
\gls{CRIU} supports incremental memory checkpointing capturing and saving only state changes across multiple intervals to minimize checkpoint size and reduce downtime \cite{singh2022ldt}.
Practical application of \gls{CRIU} including successful live migration, rapid initialization, and its integration into Kubernetes (an open-source platform for automating container orchestration and management) has been confirmed in practice~\cite{criu_application}.
Furthermore, empirical studies demonstrated the potential benefits of \gls{CRIU} in Edge computing, particularly in conjunction with Docker’s experimental features, as well as its impact on container performance~\cite{criu_application2}.
It can be concluded that \gls{CRIU} in its current state is viable for individual containers when time constraints are not a primary concern, allowing applications with flexible time requirements to benefit from its functionality without significant impediments.

\subsubsection{Process Hauler} 

Built on top of \gls{CRIU}, \gls{P.Haul} emerged to address a specific limitation of \gls{CRIU}.
While \gls{CRIU} provides the basic functionality needed for live migration, such as pre-dumping the runtime state of a process, it lacks a cohesive mechanism to manage and execute all the steps involved in live migration as a unified process.
This lack of orchestration makes it challenging to use \gls{CRIU} directly for live migration, especially for complex systems requiring multiple stages of state transfer.
\gls{P.Haul} addresses this gap by acting as a management layer over \gls{CRIU}.
It coordinates the various phases of live migration, particularly for scenarios where pre-copy migration techniques are required.
However, since \gls{P.Haul} performs image migration by synchronizing the \gls{rootfs}, when it comes to OCI\footnote{Standardized container images that adhere to the Open Container Initiative Image Specification. These images contain a set of read-only filesystem layers that form the basis for the container's runtime environment.} compatible images, it is unable to take advantage of the layered structure of images to prevent the transmission of redundant layers~\cite{sledge_paper}.
To the best of our knowledge, no significant research work has been published solely on \gls{P.Haul}.
However, OpenVZ and \gls{LXC} attempted to make use of \gls{P.Haul} \cite{Ma2017edge, Qiu2017lxc}.


\subsubsection{Fault Tolerance Interface}
Designed for \gls{HPC} environments, \gls{FTI} is a user-level checkpoint/restore library~\cite{Bautista2011}. 
It allocates host buffers for checkpoint data and thus can synergize with live migration techniques.
Coupled with resource managers, it offers transparent operation which forms a comprehensive solution for resilient and flexible \gls{HPC} environments.
Both \gls{FTI} and \gls{DMTCP} are designed for \gls{HPC} environment.
However, due to seamless integration with resource managers like Slurm, and \gls{RDMA} network compatibility, \gls{DMTCP} outperforms \gls{FTI} as a versatile solution in live migration, in particular for \gls{HPC} environments.
 
\subsubsection*{Platform Specific Tools}
In addition to the aforementioned tools, several platform specific tools are available.
\textit{Redhat OpenShift} supports live migration for both containerized and \gls{VM}-based workloads.
In terms of containers, OpenShift supports stateful application workload migration within or between OpenShift clusters via \gls{MTC} at the granularity of a namespace; meaning it migrates not just one container, but all namespace-level objects within it \cite{openshift_mtc}.
\textit{Fedora Checkpoint Restore} utilizes \gls{CRIU}, and is implemented through the crtools package and specific kernel configurations.
It enables transparent OS-level migration of processes, process trees, or containers between Fedora systems without service interruption, and does not require application modifications.
\textit{Hyper-V Checkpoint Restore} is a feature in Microsoft's virtualization platform that enables saving and restoring virtual machine states.
It supports both memory state and data-consistent captures.
Combined with live migration capabilities, Hyper-V \gls{C/R} allows for seamless transfer of running \gls{VM}s between hosts.

\subsubsection*{Emerging Tools}
Beside well-established tools, several emerging tools offer promising alternatives, in particular enabling effective migration across heterogeneous platforms.
\textit{H-Container} enables the migration of natively compiled containerized applications across compute nodes with CPUs of different \gls{ISA}s~\cite{Barbalace2020}.
It extends Popcorn Linux to a container environment \cite{Bapat2024popcorn} and is built upon \gls{CRIU}.
H-Container uses \gls{IR} lifting software to instrument applications for cross-\gls{ISA} migration without requiring the source code~\cite{Barbalace2020}.
\textit{\gls{HETCOM}} focuses on secure container migration in heterogeneous computing environments. It ensures security by using \gls{TEE}s on edge nodes and using \gls{TPM} 2.0 on cloud nodes~\cite{Wruck2024}. 
\gls{HETCOM} is not bound to any container engine and does not rely on any existing \gls{C/R} tool.
\gls{HETCOM} aims for seamless integration with emerging containerization technologies while imposing low overhead.
\textit{Wharf} introduces a new paradigm for transparent and efficient live migration across heterogeneous hosts.
Wharf introduces an abstraction called “vessel” to provide a machine-independent representation of an application's state, and “dock” which is a runtime layer for executing and migrating vessels, allowing transparent live migration between different \gls{ISA}s and operating systems.
The authors assert that Wharf's checkpoint and restore time is, on average, 14× faster than \gls{CRIU}~\cite{yangwharf2024}.

\section{Migration Unit}
\label{sec:units}
The migration unit is the computational entity transferred between hosts during a migration.
As discussed in Section~\ref{sec:overview}, in most scenarios, this unit is optimally either a container, a group of containers (pod), or a virtual machine, each requiring distinct approaches and tools for effective migration.
The selection of the migration unit affects the complexity, resource requirements, and performance characteristics of the migration process, thereby shaping the overall strategy and implementation of live migration in diverse computing environments.
The decision made here influences and is shaped by several key elements, including the volume of data to be transferred, the time required for the process, and system compatibility.

\subsection{Container}
In the modern computing landscape, containerization has revolutionized software application development, distribution, and deployment.
This approach involves packaging an application with its essential components, including binary code, libraries, dependencies, and configuration files into lightweight, portable, and consistent environments known as containers~\cite{Gamess2024} which generally incur less virtualization overhead compared to virtual machines~\cite{Govindaraj2018}.

The container landscape primarily consists of application containers and system containers.
Application containers are lightweight, standalone packages that encapsulate a single application and its dependencies,
while system containers are more comprehensive units and include a full operating system environment, allowing them to run multiple applications or services within a shared kernel space.
Both types of containers share the host operating system's kernel.
System containers, in addition, provide an isolated user space, offering a complete environment.

In the context of containerization, a container engine typically provides higher-level functionality, including image management, build processes, and user interfaces.
The container runtime, on the other hand, focuses on the core tasks of creating and running containers.
Given the rapid development and adoption of containerization concept in research and industry, the terminology isn't always consistent, and terms are often used interchangeably or with slightly different meanings depending on the context.
Therefore, this work uses “container technology” to refer to both container engine and runtime, unless a specific distinction is needed.
Table~\ref{tab:containerization-technology} provides a list of container technologies categorized based on their main functionality.

Among the largely adopted container technologies, Docker and Podman primarily focus on application containers, while \gls{LXC} is designed for system containers~\cite{analysis_docker_lxd}, and Apptainer targets \gls{HPC} environments.
A growing standardization effort aims to improve interoperability among container technologies, enabling interchangeable components across frameworks and better compatibility across computing systems.
Consequently, the boundaries between existing technologies are often blurred, making it difficult to draw clear distinctions in many cases.

\begin{table}[h]
\caption{Containerization and Orchestration Technologies}\label{tab:containerization-technology}
\begin{tabular*}{\textwidth}{@{\extracolsep\fill}lll}
\toprule
Category & Technology & Description \\
\midrule
\rule{0pt}{2.5ex}\multirow{4}{*}{Engine}                          & Docker                         & Daemon-based                       \\
                                                                  & Podman                         & Daemonless                         \\
                                                                  & \gls{LXC}                      & OS-level                           \\
                                                                  & Apptainer (aka Singularity)    & \gls{HPC}-focused                  \\ \hline
\rule{0pt}{2.5ex}\multirow{3}{*}{Runtime}                         & Containerd                     & Low-level                          \\
                                                                  & CRI-O                          & for Kubernetes                     \\
                                                                  & runc                           & Low-level                          \\ \hline
\rule{0pt}{2.5ex}\multirow{2}{*}{Self-hosted Orchestration}       & \gls{K8s}                      & Large-scale                        \\
                                                                  & Docker Swarm                   & Small-scale                        \\ \hline
\rule{0pt}{2.5ex}\multirow{4}{*}{ Managed Orchestration Service}  & Google \gls{K8s} Engine (GKE)  & Managed \gls{K8s} by Google Cloud  \\
                                                                  & Amazon \gls{K8s} Service (EKS) & Managed \gls{K8s} by AWS           \\
                                                                  & Azure \gls{K8s} Service (AKS)  & Managed \gls{K8s} by MS Azure      \\
                                                                  & Oracle \gls{K8s} Engine (OKE)  & Managed \gls{K8s} by Oracle Cloud  \\ \hline
\rule{0pt}{2.5ex}\multirow{3}{*}{Specialized}                     & RedHat OpenShift               & Enterprise Kubernetes              \\
                                                                  & MicroK8s                       & Kubernetes for IoT                 \\  
                                                                  & K3s                            & Kubernetes for IoT                 \\
\botrule
\end{tabular*}
\end{table}

In the following section, the most relevant container engines with respect to their support for live migration are discussed.

\subsubsection{Linux Container} 
A system environment close to a full Linux system is provided by \gls{LXC} by launching an OS in every container.~\cite{survey_live_migration_container,analysis_docker_lxd,lxc_documentation}. 
\gls{LXC} leverages the namespace mechanism to isolate containers and utilizes native Linux \gls{CGroups} to share the kernel with containers while managing CPU, memory, disk and network resources.
The concept of \gls{LXC} makes it in many directions comparable to \gls{VM}s~\cite{Kaur2022cloud}.
\gls{LXC} integrates \gls{CRIU}, however, it does not yet support checkpointing of network namespaces~\cite{Rohit2023edge}.
An enhanced and user-friendlier version of \gls{LXC} is \gls{LXD} that is built around REST APIs to communicate with \gls{LXC} software library (liblxc), enhancing the \gls{LXC} features such as improved security and isolation, live migration and limiting resource of the containers~\cite{ Qiu2017lxc, analysis_docker_lxd, containerization_book}.

\subsubsection{Docker}
\label{subsubsec:docker}
Initially built on \gls{LXC}, Docker later transitioned to its own container runtime library, libcontainer, to achieve greater control and flexibility.~\cite{analysis_docker_lxd}.
Studies show that on average less than $7\%$ of the content inside a Docker image is used during the runtime, thus, the layered property of Docker images has been leveraged to further reduce the migration overhead~\cite{Harter2016docker}.
A similar study found that each Docker image layer has a unique SHA256 local ID and proposed a method to prevent redownloading layers with different SHA256 IDs. Other works such as Sledge~\cite{sledge_paper}-- a live migration system, are focused on fine-tuning the framework by introducing \gls{LCR} for container images, utilizing \gls{CRIU}'s incremental memory checkpointing ability for an improved runtime, and by designing \gls{DCL} scheme for effective management.
Voyager~\cite{Nadgowda2017Voyager}, a just-in-time migration tool built on \gls{CRIU}, that minimizes downtime by employing data federation via union mounts.
It utilizes tmpfs for dump files to avoid storage latencies~\cite{Cao2024edge} and adopts a multi-pronged migration approach: \gls{CRIU} handles in-memory state transfer, a dual-band system manages local filesystem migration between source and target, and network filesystems are migrated by swapping mount points.
This tool is highly dependent on network-attached storage and upon any degradation of network quality, the migrated service may experience frequent errors in the background-replicated files, leading to a general performance degradation~\cite{Shubha2025CSMD}.
Other research efforts focus on minimizing data transmission by optimizing Docker's storage mechanisms and employing algorithms such as \gls{JNLP} to reduce dirty memory pages~\cite{Kang2023docker}.
Experimental results demonstrate decreased migration times for both stateless and stateful containers.

\subsubsection{Podman} 
As a daemonless, Linux-native, open-source tool developed by RedHat, Podman is built on runc and allows users to create OCI-compliant containers.
It is based on libpod, a library for container lifecycle management and provides APIs for managing containers, pods, container images, and volumes~\cite{github_libpod}.
Podman's daemonless design allows containers to run rootless by default, eliminating the need for background services such as Docker’s dockerd.
Podman also integrates \gls{CRIU} with pre-copy optimization, enabling checkpointing and restoring containers across hosts. 
Although based on our practical experience, stateful checkpointing and restoring of Podman containers works smoothly on RedHat-based distributions, containers relying on a systemd-based entry point may encounter issues during checkpointing since \gls{CRIU} cannot fully capture systemd's state, and active TCP connections require the \textit{-{}-tcp-established} flag for the connections to be properly handled.
Several studies also note that performing reliable live migration via Podman requires root permission~\cite{github_libpod,podman-doc}. 
Except for Podman, which requires \gls{CRIU} to skip network namespace recreation while restoring a checkpointed container, other container technologies let \gls{CRIU} handle namespaces~\cite{reber2019,podman-doc}.

\subsubsection{Apptainer}
Previously known as Singularity, it was originally developed for \gls{HPC} ecosystems.
Apptainer container engine offers features such as native support for GPUs, MPI, InfiniBand, and resource managers like Slurm~\cite{apptainer-doc}.
Unlike Docker and \gls{LXC}, it does not rely on \gls{CGroups} which means less strict resource constraints, and operates without root daemon.
In contrast to Docker and Podman, Apptainer does not provide built-in support for \gls{CRIU}~\cite{Fabio2024criu}.
One reason is the single-file image format of Apptainer in contrast to the layered file systems, which makes it an undesirable candidate for the incremental checkpointing feature of \gls{CRIU}.
Besides that, Apptainer bypasses the host library stack, preventing the preloading of libraries required to enable checkpointing.
There has been various efforts to enable checkpointing for Apptainer containers using \gls{DMTCP}~\cite{sing-dmtcp,sing-dmtcp2}.
In addition, the required checkpointing libraries can be also manually injected via an specfile to realize a checkpointable Apptainer container.
However, according to Apptainer's documentation, checkpoint/restart functionality remains on the roadmap.



\subsection{Virtual Machine}
As a well-established technology, \gls{VM} live migration enables the seamless transfer of running \gls{VM}s between physical hosts without service interruption.
Despite challenges like managing large memory states and network connections, the technology has evolved to enhance resource utilization and system reliability.
As virtualization continues to advance, \gls{VM} live migration capabilities are becoming increasingly efficient, addressing more complex scenarios and larger \gls{VM} instances.
Table~\ref{tab:vm-technology} provides a list of \gls{VM} technologies categorized based on their main functionality.

Majority of production \gls{VM} live migration implementations rely on hypervisors, which utilize pre-copy and post-copy techniques.
Hypervisor provides hardware virtualization capabilities and is categorized into two types.
Type 1 hypervisors, also known as bare-metal hypervisors, interact directly with the host machine's physical hardware, bypassing the need for an underlying operating system.
This direct access allows for more efficient resource allocation and management, resulting in optimal performance and reduced overhead.
In contrast, Type 2 hypervisors, also called hosted hypervisors, run as applications within an existing operating system.
They access hardware resources through the host OS, which can lead to lower performance compared to Type 1 hypervisors due to the additional software layer.

\begin{table}[h]
\caption{Virtual machine technologies}\label{tab:vm-technology}
\begin{tabular*}{\textwidth}{@{\extracolsep\fill}lll}
\toprule%
Category & Technology & Description \\
\midrule

\rule{0pt}{2.5ex}\multirow{4}{*}{Type 1 Hypervisor} & VMware vSphere             & Enterprise-grade                 \\
                                                    & Microsoft Hyper-V          & Windows Server                   \\
                                                    & Xen                        & Open-source                      \\
                                                    & \gls{KVM}                  & Kernel-based                     \\ \hline
\rule{0pt}{2.5ex}\multirow{3}{*}{Type 2 Hypervisor} & Oracle VirtualBox          & Free, cross-platform             \\
                                                    & VMware Workstation         & Desktop virtualization           \\
                                                    & Parallels Desktop          & Mac-focused                      \\ \hline
\rule{0pt}{2.5ex}\multirow{4}{*}{Cloud based}       & Amazon EC2                 & AWS \gls{VM}s                    \\
                                                    & Google Compute Engine      & GCP \gls{VM}s                    \\
                                                    & Microsoft Azure VMs        & Azure \gls{VM}s                  \\
                                                    & OpenStack                  & Open-source                      \\ \hline
\rule{0pt}{2.5ex}\multirow{3}{*}{Management}        & VMware vCenter             & Centralized                      \\
                                                    & Microsoft \gls{VM}Manager  & For Hyper-V                      \\
                                                    & Citrix XenCenter           & For XenServer                    \\ \hline
\rule{0pt}{2.5ex}Emulators                          & \gls{QEMU}                 & Open-source                      \\ \hline
\rule{0pt}{2.5ex}\multirow{4}{*}{Specialized}       & Wind River Hypervisor      & Embedded systems                 \\
                                                    & User Mode Linux            & Linux kernel-based               \\
                                                    & RedHat OpenShift           & Container and \gls{VM}           \\
                                                    & Harvester                  & Open-source, HCI            \\  
                                   
\botrule
\end{tabular*}
\end{table}

\subsubsection{VMware vSphere}
The vSphere platform, with its Type 1 ESXi hypervisor and vCenter Server for management, supports application virtualization, high availability, and disaster recovery, while optimizing performance, security, and deployment~\cite{vmware_vsphere_book}.
VMware vSphere vMotion enables live migration of workloads between hosts. 
Unlike vSphere \gls{HA}, which automatically restarts \gls{VM}s on other servers after a failure, vMotion focuses on workload balancing and maintenance without requiring \gls{VM} restarts.
Recent studies show that machine learning approaches can effectively improve the overall performance of vMotion in live migration scenarios~\cite{elsaid2022}.


\subsubsection{Microsoft Hyper-V}
As a Type 1 hypervisor, Hyper-V transfers running virtual machines between physical hosts. 
Hyper-V works similarly to VMware vMotion, with best practice recommending a dedicated \gls{VLAN} to isolate migration traffic from application traffic~\cite{live_migration_hyper_v,live_migration_hyper_v2}.

\subsubsection{Xen}
As an open-source Type 1 hypervisor, Xen supports \gls{VM} migration between different Xen-powered servers and enables the execution of multiple operating systems on the same machine. 
Xen uses the pre-copy mechanism by default, to provide live migration on all versions of XenServer~\cite{live_migration_cloud_centre}.
Migration of Xen \gls{VM}s to \gls{KVM}  is also possible for various Linux distributions.
Since XenServer 6.1, storage and block migration techniques are also supported which enables live migration of \gls{VM}s without shared storage.
Xen supports two virtualization types: \gls{PV}, where guest OSes interact directly with the hypervisor for resources, and \gls{HVM}, where unmodified OSes run with virtual hardware emulation, resulting in higher overhead.~\cite{containerization_book, xen_beginner_guide}.



\subsubsection{Kernel-based Virtual Machine}
\gls{KVM}, a Linux kernel module enabling hypervisor functionality, has been part of the official kernel since version 2.6.20.
\gls{KVM} supports live migration, 
while the \gls{VM} stays powered on, maintains network connectivity, and keeps applications running during relocation.
It also enables checkpointing a \gls{VM}'s state for later resumption.

\subsubsection{Quick Emulator} 
\gls{QEMU} can emulate an entire system, including the processor and various peripherals. It supports live migration using the pre-copy approach, where the VM’s memory is transferred in multiple iterations while it continues running, followed by a brief pause to send the final dirty pages and CPU/device state, and post-copy approach, where the memory migrates upon the destination
VM’s \gls{NPT} faults \cite{STORNIOLO2024, jian2025}. To reduce migration time and downtime in pre-copy, \gls{multifd} enables parallel transfer of memory pages using multiple threads \cite{yang2024, jian2025}. The new mapped-ram stream format complements this by assigning fixed offsets to RAM pages in the migration file, allowing concurrent, conflict-free writes and enabling efficient I/O with O\_DIRECT (write directly to disk and skip caching) \cite{qemudoc}. 




\subsubsection{OpenStack}
As an open-source cloud platform, OpenStack manages compute, storage, and networking resources within a data center.
It also allows users to perform live migration of \gls{VM}s on \gls{KVM}-libvirt and Xen hypervisors, provided that certain conditions are met across all nodes, such as having a shared instance path across compute hosts, enabling passwordless SSH for root access, and configuring firewalls to allow libvirt communication.
If the rate of dirty memory pages is high during live migration, it may fail, but workarounds such as post-copy migration or auto-converge mechanism, which throttles the source worker's CPU to slow it down and keep up with memory changes, are proposed~\cite{Mirkhanzadeh2021converge,openstack_livemigration}.
When using \gls{KVM}-libvirt, these limitations must be considered, whereas for Xen, different requirements apply, including the use of compatible XenServer hypervisors and an NFS export visible to all XenServer hosts~\cite{openstack_livemigration}.

\subsection*{Hybrid and Emerging Solutions}

As listed in Table~\ref{tab:containerization-technology} and~\ref{tab:vm-technology}, the virtualization landscape extends beyond these well-established open-source technologies and includes a diverse range of solutions.
OpenVZ, primarily a container technology, offers features that share similarities with both containers and \gls{VM}s~\cite{Walters2008}. 
OpenVZ has its own set of utilities for managing containers, such as vzctl and prlctl~\cite{di2020reducing, containerization_book}.
In 2006, OpenVZ introduced its own live migration approach for containers, which is also compatible with VMs.
By treating containers as regular processes, OpenVZ reduces management overhead.
Comparative studies of container live migration capabilities between OpenVZ and \gls{LXC} have shown mixed results.
While \gls{LXC} demonstrated shorter total migration times and downtimes, OpenVZ exhibited a less intrusive approach with lower resource utilization during the live migration process.
This highlights OpenVZ's efficiency in resource management, despite slightly longer migration durations~\cite{indukuri2016performance}.

Oracle VirtualBox is another open-source, cross-platform virtualization software that
supports live migration, called “Teleporting”.
While VirtualBox allows for cross-platform live migrations, its live migration is generally considered less advanced than other virtualization platforms like Hyper-V.

Evolving \gls{HCI} platforms like the open-source Harvester, built on Kubernetes, now run both \gls{VM}s and containers on bare metal, integrating KubeVirt for virtualization and Longhorn \cite{kampadais2025longhorn} for distributed storage.
Nevertheless, live migration in these technologies still faces challenges.
For instance, migrating virtual machines between nodes with different CPU models remains problematic in certain scenarios~\cite{Harvester}.

Built based on \gls{KVM}~\cite{habib2008kvm} and KubeVirt~\cite{Karlsson2024kubeVirt}, RedHat OpenShift also supports live migration for both containers and \gls{VM}s using shared storage with \gls{RWX} access mode, blurring the lines between these technologies and demonstrating overlap in management platforms.
Leveraging the \gls{MTV}, it enables the migration of \gls{VM}s from other hypervisors (VMware vSphere, \gls{RHV}, and OpenStack) into the OpenShift environment~\cite{openshift_mtc}.

This convergence within a single platform highlights the evolving cloud-native landscape, where the distinction between virtualization methods is becoming less defined.
Major technology companies including Google, Amazon, Oracle, and Microsoft also offer proprietary solutions in the form of \gls{IaaS} that address specific enterprise needs.
These technologies span cloud computing, containerization, and hybrid environments.
Furthermore, specialized technologies are emerging to extend migration capabilities to Edge devices, facilitating new applications in \gls{IoT} and Edge computing scenarios.
This diversification reflects the ongoing innovation in virtualization technologies, driven by increasing demands for scalability, performance, and flexibility across various computing environments.
Table~\ref{tab:checkpoint-comparison} presents a comparative overview of the key features of prominent containerization and virtualization technologies.

As shown in Table~\ref{tab:checkpoint-comparison} while process-level solutions such as \gls{CRIU} and \gls{DMTCP} offer fine-grained migration capabilities, they are limited in user transparency, platform integration, and support for distributed or \gls{HPC} workloads, with \gls{DMTCP} providing stronger support for distributed applications at the cost of increased overhead.
OpenVZ presents a balanced approach with low overhead and good user transparency but suffers from limited modern platform integration due to its reliance on specialized kernels.
In contrast, hypervisor-based solutions like VMware vMotion and KVM/QEMU stand out for their robust, seamless, and highly transparent live migration at the virtual machine level, offering minimal overhead, broad application compatibility, and strong integration with contemporary cloud and enterprise environments.
Ultimately, the choice of technology should be guided by the required migration granularity, application type, and operational environment, with \gls{VM}-level solutions proving most effective for general-purpose and cloud workloads, and process-level tools remaining relevant for specialized scenarios such as \gls{HPC} and tightly-coupled distributed systems.


\begin{sidewaystable}[htbp]
\caption{Comparative overview of live migration support by prominent technologies}\label{tab:checkpoint-comparison}%
\begin{tabular}{@{}llllll@{}}
\toprule
Technologies & \gls{CRIU}-based\footnotemark[1]  & \gls{HPC}/\gls{DMTCP}-based\footnotemark[2] & OpenVZ &  VMware vMotion & \gls{KVM}/\gls{QEMU}\\
\midrule
Migration unit           & Process or Container      & Process (across machines)        & Full \gls{VE}state   & \gls{VM}                & \gls{VM}       \\
Implementation           & Kernel level(ptrace API)  & Library level(intercepts calls)  & Kernel extension     & Hypervisor-based        & Hypervisor-based \\
Prerequisite             & Recent kernels            & Special launch script            & OpenVZ kernel        & VMware                  & \gls{KVM} support  \\
Userspace vs. Kernel     & Requires kernel modules   & Runs in userspace                & Kernel-based         & Hypervisor-based        & Hypervisor-based  \\
Platform integration     & Docker and Kubernetes     & No                               & No                   & No                      & Yes     \\
App compatibility        & Wide range                & Limited                          & Wide range           & Wide range              & Wide range   \\
App transparency         & Yes                       & Yes                              & Yes                  & Yes                     & Yes     \\
User transparency        & No                        & Partial                          & Yes                  & Yes                     & Yes             \\
Distributed Apps         & Limited                   & Strong support                   & Across hosts         & Long-distance           & Across hosts    \\
\gls{HPC} relevance      & Limited                   & Yes                              & Yes                  & Yes                     & Yes        \\
Overhead                 & Generally lower           & Issues due to proxying           & Minimal              & Minimal                 & During migration  \\
PID handling             & Maintains original PIDs   & May use fake PIDs          & Maintains PIDs in \gls{VE} & N/A (\gls{VM}-level)    & N/A (\gls{VM}-level)   \\
Net connection           & No open connection        & Yes                              & Yes                  & Yes                     & Yes        \\
\botrule
\end{tabular}
\footnotetext[1]{The \gls{CRIU}-based column includes technologies such as \gls{P.Haul}, which are based on \gls{CRIU}.}
\footnotetext[2]{The \gls{DMTCP}-based column refers to solutions, which are based on \gls{DMTCP}.}
\end{sidewaystable}

\section{Infrastructure Characteristics}
\label{sec:characteristics}
Infrastructure characteristics, including network conditions and resource availability, directly affect live migration's performance, efficiency, and reliability.
These characteristics, combined with environmental preferences and constraints, influence the overall strategy for live migration.

\subsection{Network Conditions}
The process of migration involves substantial data transfer, making network conditions crucial for its success.
Particularly when transitioning from \gls{LAN}s to \gls{WAN}s, live migration presents significant challenges.
Key network-related factors include bandwidth, latency and congestion, and IP reassignment.
\gls{WAN}s typically offer lower bandwidth compared to \gls{LAN}s, resulting in extended migration times and increased downtime, especially when transferring large volumes of storage data (e.g., in shared nothing setups).
Unlike \gls{LAN}s, migrations over \gls{WAN} necessitate IP reassignment, leading to reconfiguration issues.
The risk of data loss due to network failures increases during \gls{WAN} migration, although techniques such as replication and erasure coding help mitigate this risk.
When migrating multiple virtual machines simultaneously, network resources become strained, requiring efficient scheduling and bandwidth utilization to minimize service disruptions.
These challenges collectively underscore the complexity of \gls{WAN}-based live migration and highlight the need for careful planning and optimization strategies~\cite{Raseena2022network, choudhary2017network}.

\subsection{Resource Availability}
Sufficient resources such as CPU/GPU and memory are essential for efficient execution of migration tasks, and maintaining performance of both source and target hosts during migration~\cite{widjajarto2021live}.
Adequate storage capacity and performance are also crucial for accommodating migrated data, and for ensuring smooth transfer of large datasets~\cite{Govindaraj2018}.
Compatibility between software and hardware stacks on both source and target hosts as well as timely allocation of required resources impose further challenges and influence live migration strategies.

\subsection{Cluster Environment}
Clusters, in classic terms, are a type of computing where multiple computers are connected to work together as a single system.
They are typically designed to provide high computational power and are often used for specific tasks or workloads that require significant processing capabilities.
A prime example of clusters is \gls{HPC} clusters, which aggregate computing resources to tackle complex, large-scale workloads.
In \gls{HPC} clusters particularly, GPUs play a crucial role. 
Although live migration for GPU-enabled \gls{VM}s is possible, since GPUs maintain extensive internal states, it is significantly more challenging compared to CPU migration, and highly depends on the specific technology and setup~\cite{rapture2022}.
NVIDIA vGPU supports live migration via mediated passthrough when using NVIDIA GPUs with vGPU technology~\cite{xu2019characterization}.
Hyper-V in Windows Server 2025 supports live migration with \gls{GPU-P}, although there are reported difficulties in its implementation.
gMig reported successful live migration for full virtualization on Intel GVT-g GPUs with an average downtime of 302ms on Windows and 119ms on Linux operating systems~\cite{tian2014full, ma2018gMig, Belkhiri2024GPU}.
However, it only supports Intel GPUs, and the migration process must wait for GPU tasks to complete before migration, making this approach impractical for long-running GPU workloads.
\gls{POS} is a system designed to improve GPU checkpointing and restoring by enabling concurrent execution of \gls{C/R} with GPU tasks~\cite{Zhuobin2024GPU}.
However, periodic checkpointing of GPU memory can lead to significant overhead as examined by other studies~\cite{Mingcong2024GPU}.
\gls{NVCR} is an extension for system-level checkpoint software like \gls{BLCR} and \gls{DMTCP}, enabling transparent checkpointing and restarting of CUDA applications without requiring modifications from developers~\cite{Nukada2023cuda}.
As mentioned earlier, \gls{CRIU} although being the de-facto choice for \gls{C/R}, struggles with tasks involving device states managed by drivers.
While \gls{CRIU} offers a plugin mechanism to address this limitation, there are currently no publicly available plugins for complex devices like GPUs.
AMD plans to add ROCm support to \gls{CRIU} by creating plugins that save and restore GPU state, enabling checkpointing and restoring of ROCm-based applications~\cite{AMDCRIUPlugin2025}.

\subsection{Cloud Environment}
Cloud computing, primarily focuses on providing users with on-demand access to a shared pool of configurable computing resources. 
It emphasizes flexibility, scalability, and the ability to provision resources rapidly\cite{Attar2017}.
Although traditionally, clusters and cloud computing have had distinct characteristics and purposes, the distinction between clusters and clouds has become less clear-cut in recent years~\cite{munhoz2024HPCCloud, Sherpa2024}.
Cloud providers now offer \gls{HPC} solutions that resemble traditional clusters.
In cloud computing environments, specific features such as multi-tenancy and elasticity impact migration.
Shared infrastructure among multiple tenants necessitates careful resource management during migration to maintain isolation and performance~\cite{survey_live_migration_container}.
The ability to scale resources dynamically can facilitate more efficient migrations by allocating additional resources as needed~\cite{live_migration_data_management}.
However, the trade-off between minimizing migration time and ensuring acceptable quality of service is challenging~\cite{Breitgand2010cloud}.
Cloud environments due to their vast amount of resource availability can benefit from certain approaches such as resource reservation for accelerated migration~\cite{Gupta2022Cloud, Ye2011vm}.
Although, modern cloud platforms provide support for containers, due to the maturity of \gls{VM} technology in comparison to containerization, most studies on the topic of live migration consider cloud environments solely consisting of \gls{VM}s~\cite{Breitgand2010cloud, Gupta2022Cloud,Ye2011vm}.
There are only a few recent studies such as a three-tiered kubernetes-based cloud ecosystem performing migration of \gls{CNF} which focuses on live migration of containers in cloud environment~\cite{Kaur2023kubernetes}.
However, in this approach, a pod is terminated and then recreated, causing the loss of container state.
To address this, preserving container states on a persistent volume, and injecting the state into the new pod upon recreation was proposed, which in turn also reduces the startup time~\cite{Vasireddy2024kubernetes}.
Kubernetes recently introduced \gls{C/R} support with CRI-O as the container runtime, promoting it to Beta in version 1.30. It is accessible via kubelet APIs, not kubectl, requiring execution directly on the node where kubelet runs.
So despite its progress, the field remains in its early stages and necessitates additional research.

\subsection{Edge and Fog Environment}
Edge computing and Fog computing are both distributed computing paradigms that aim to bring computation and data processing closer to the source of data generation~\cite{Govindaraj2018}.
While Edge computing is more decentralized and localized, Fog computing creates a hierarchical structure that facilitates data aggregation and preprocessing.
Although Fog computing is often used interchangeably with Edge computing, especially in smaller setups; in larger deployments, Fog computing serves as a distinct layer between Edge computing and cloud computing~\cite{edge_computing2}.
The distinguishing characteristics of Edge computing in comparison to cloud and cluster computing are geographical distribution, mobility support, location awareness, proximity, low latency and heterogeneity~\cite{edge_computing_survey}.
Since latency is critical in Edge computing, most live migration research in this field focuses on minimizing downtime, migration time, and overall latency.
In addition, Edge computing is closely tied to resource-constrained servers, which makes live migration primarily container-based.
In contrast, Fog computing, with its resource-rich infrastructure, often involves \gls{VM} migration~\cite{attar2021}.
Several recent studies have explored live migration techniques in Fog and Edge computing environments.
Such as proposing an approach using reinforcement learning for decision-making in live \gls{VM} migration, considering multiple factors to address early and late handover issues~\cite{alqam2023reinforcement}, or introducing $EVM\_MIG$ system that considers device type, function, and location for \gls{VM} migration decisions in heterogeneous Fog environments~\cite{alqam2023location}.


Furthermore, \gls{LIMOCE}, was proposed as a lightweight algorithm for live migration in Edge computing~\cite{Rohit2023edge}.
Other studies proposed smart pre-copy live migration approach for \gls{VM} migration, aiming to minimize downtime and migration time~\cite{Osanaiye2017FromCT}, or enhancement to the post-copy live migration algorithm to make it robust against failures in Fog environments~\cite{talebian2022robust}.

\section{Discussion and Conclusion}
\label{sec:discussion}
Live migration offers numerous benefits but requires a well-planned strategy.
While environment characteristics guide the choice of live migration unit and technique, the ultimate decision depends on migration \textit{objectives} like service availability or energy efficiency, which may sometimes conflict or even discourage live migration altogether.

Across the virtualization spectrum, bare-metal live migration remains impractical, particularly in heterogeneous environments.
Conversely, \gls{VM} live migration technologies have achieved maturity and widespread adoption, offering reliable performance despite higher overheads.
Container live migration occupies an intermediate position, providing enhanced speed and reduced overhead compared to \gls{VM}s while accommodating diverse environments.
Although rapidly evolving and approaching production readiness, container migration continues to face challenges in complex scenarios and is still gaining broader acceptance, especially for latency-sensitive applications.
Table~\ref{tab:checkpoint-comparison} offers a general framework for planning live migration strategies.
Additionally, several key factors should be taken into consideration.
GPU live migration remains a significant challenge for both virtual machines and containers.
In scenarios necessitating the preservation of external network bindings, virtual machines generally offer superior capabilities compared to containers.
It is strongly advised to utilize recent Linux kernel versions, as leading techniques depend on \gls{CGroups} functionality and recent system calls.
Older kernels lack support for crucial features, such as maintaining \gls{PID}s, which are essential for effective implementation of live migration.
Most container live migration techniques, relying on \gls{CRIU}, share common limitations, including inconsistent checkpointing of full process states in complex setups.
Using techniques that support OCI-compatible images is recommended, as their layered structure enables efficient incremental migration.
In critical environments, live migration may be impractical due to the requirement of root access for most migration techniques. 
Additionally, on-demand live migration may not be feasible under specific conditions, such as network congestion or when process states are excessively large.
It is noteworthy that recent advancements in hardware technologies, particularly \gls{RDMA}, substantially enhance the broader adoption of live migration techniques by complementing existing software infrastructures, which is particularly beneficial for data-intensive tasks like deep learning, data analytics, and large-scale data transfers.

Container live migration is expected to advance with enhanced hardware support, improved performance for data-intensive tasks, and better integration with cloud-native technologies.
Integrating automated DevOps workflows with a data management system and standardized metadata schema enhances data reproducibility, aligns with FAIR\footnote{FAIR stands for Findable, Accessible, Interoperable and Reusable} principles, and adds value by documenting the migration process, data state, dependencies, and Persistent Identifiers~\cite{dsj_FAir}.
Additionally, developments will likely focus on security, Edge-to-cloud continuum support, automated migration decision-making, and improved handling of stateful applications, further expanding its applicability across diverse computing environments.

\section*{Funding}
This work was supported by the Federal Ministry of Education and Research of Germany and by the Sächsische Staatsministerium für Wissenschaft, Kultur und Tourismus in the programme Center of Excellence for AI-research “Center for Scalable Data Analytics and Artificial Intelligence Dresden/Leipzig”, project identification number: ScaDS.AI; and by the German Research Foundation (DFG) project NFDI4DataScience (no. 460234259). The funding bodies had no role in the design of the study, collection, analysis, and interpretation of data, or in writing the manuscript.


\newpage 
\printglossaries
\newpage
\bibliography{references}


\end{document}